\documentstyle[11pt,newpasp,twoside]{article}
\begin{document}
\title{Gas Dynamics and Inflow in gas-rich Galaxy Mergers}
\author{Thorsten Naab \& Andreas Burkert}
\affil{Max-Planck-Institut f\"ur Astronomie, K\"onigstuhl 17, 69117
 Heidelberg, Germany}

\begin{abstract}
We performed N-body/SPH simulations of merging gas-rich disk galaxies
with mass ratios of 1:1 and 3:1. A stellar disk and bulge component
and a dark halo was realized with collisionless particles, the gas was
represented by SPH particles. Since we did not include star formation
we focused on the gas dynamics and its influence on the structure of
merger remnants. We find that equal-mass mergers are in general more
effective in driving gas to the center. Around 50\% of the gas resides
in the central regions after the merger is complete. The gas in
unequal-mass mergers keeps a large amount of its initial angular
momentum and settles in a large scale disk while only 20\% -30\% of
the gas is driven to the center. This general result is almost
independent of the merger geometry. The gas in the outer regions
accumulates in dense knots within tidal tails which could lead to the
formation of open clusters or dwarf satellites. Later on, the gas
knots loose angular momentum by dynamical friction and successively
sink to the center of the remnant thereby increasing the total gas
content of the disk. Due to the influence of the gas the simulated
merger remnant becomes more oblate than its pure stellar counterpart.
The shape of the stellar LOSVD changes and the third-order Gauss-Hermite
coefficient $h_3$ is in  good agreement with observations.   
\end{abstract}

\section{The model}
The disk-galaxies are constructed in dynamical equilibrium (Hernquist
1993) and consist of an exponential stellar disk, a bulge with a
Hernquist profile and a pseudoisothermal dark halo. The smaller galaxy   
has 1/3 of the mass and 1/3 of the particles of the massive one. The
scale length is reduced by $\sqrt{1/3}$. In total we used 88888
particles for the collisionless component and 26666 gas particles
simulated with SPH and an isothermal equation of state. Both galaxies
approach each other on parabolic orbits with slightly inclined disks
relative to the orbital plane. All simulations were performed on a Sun
ULTRA 60 workstation. 

\begin{figure*}
\includegraphics{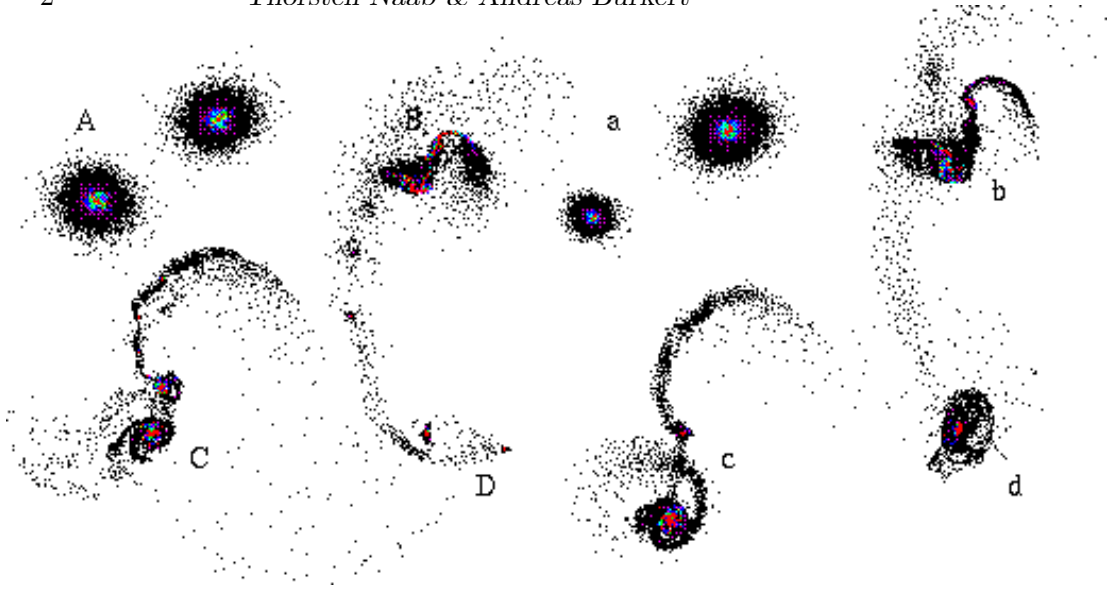}
\vspace{6cm}
\caption{Snapshots of the distribution of gas particles at different
time steps for the 1:1 (A,B,C,D) and 3:1 (a,b,c,d) merger
simulation. At the end, for the 1:1  mergers the gas is centrally
concentrated (D) whereas for the 3:1 merger the gas forms a central
disk (d)}   
\end{figure*}

\begin{figure*}
\includegraphics{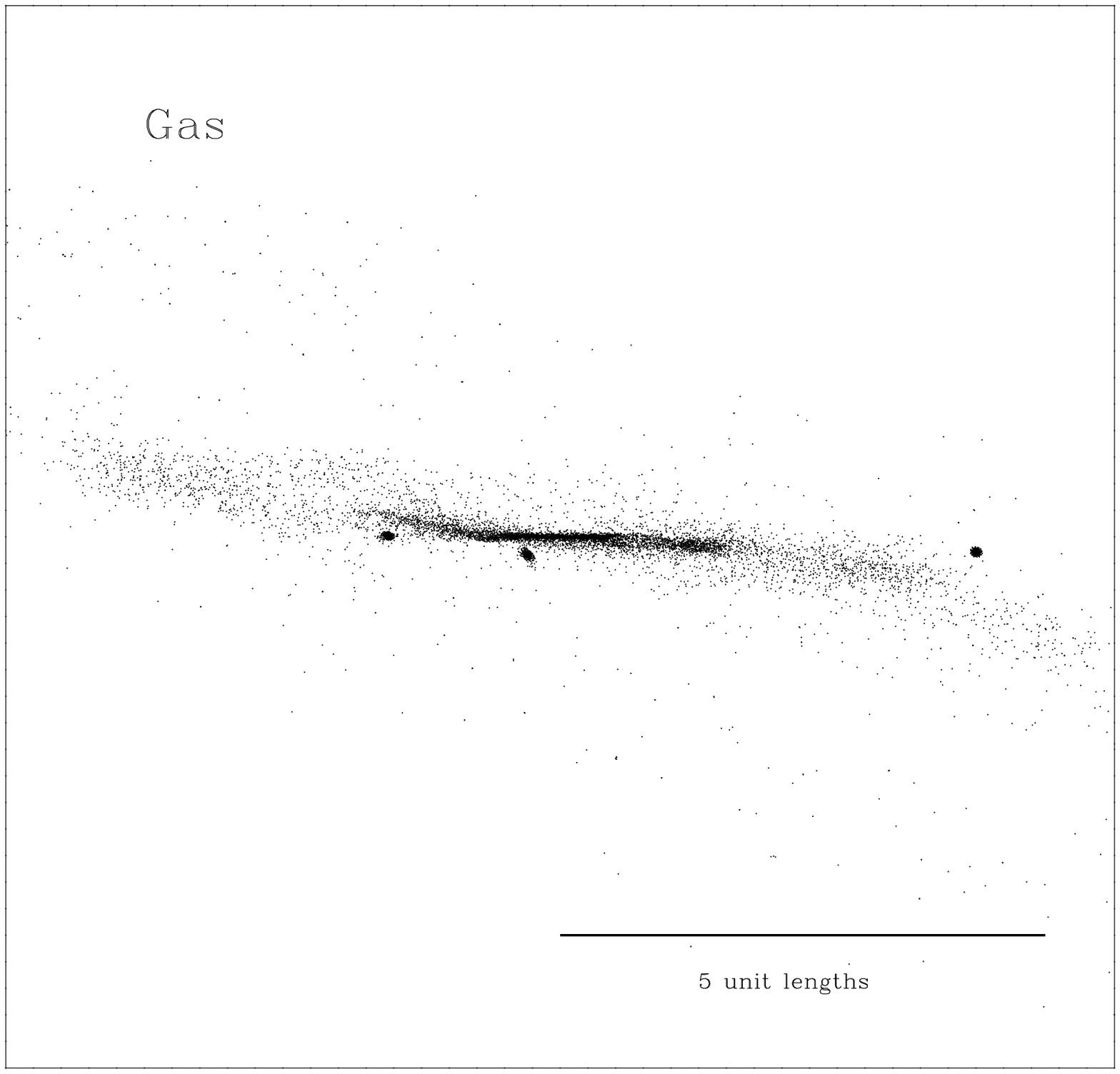}
\includegraphics{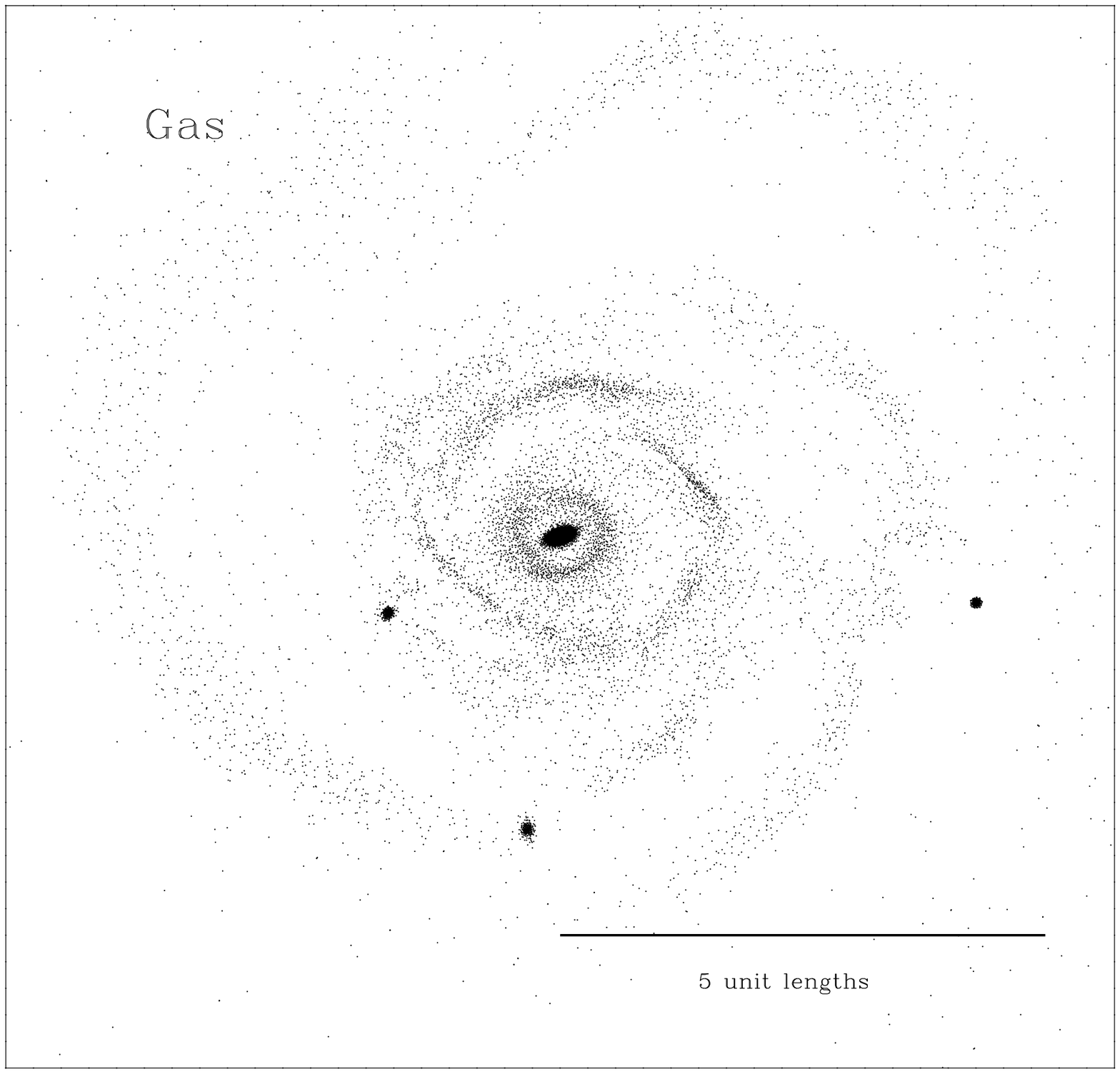}
\vspace{6cm}
\caption{The distribution of the gas particles at the and of the
simulation. In the edge on view ({\it left}) the tilt of the gas
disk is clearly visible. The face-on view is shown on the left. }
\end{figure*}

\section{Results}

Recent simulations have shown that pure stellar 3:1 mergers of
disk-galaxies lead to fast rotating (Barnes, 1998) and isotropic,
disky galaxies whereas 1:1 mergers result in anisotropic, boxy
ellipticals (Naab, Burkert \& Hernquist 1999). We tested this
hypothesis with a model that also follows the evolution of a massive
gas component (see Figure 1). During the mergers a large amount of gas
falls to the central region of the remnants (inside 300pc). Part of
the gas is driven to the center by the formation of a tidally induced
bar during the first interaction of the two disks. The
accretion rate in all cases peaks when the two galaxy 
centers merge (Figure 3). There the gas could experience a central starburst or
fuel a central black hole. Equal-mass mergers are in general more
effective in driving gas to the center (Barnes \& Hernquist 1996;
Mihos \& Hernquist 1996). \\

\begin{figure*}
\includegraphics{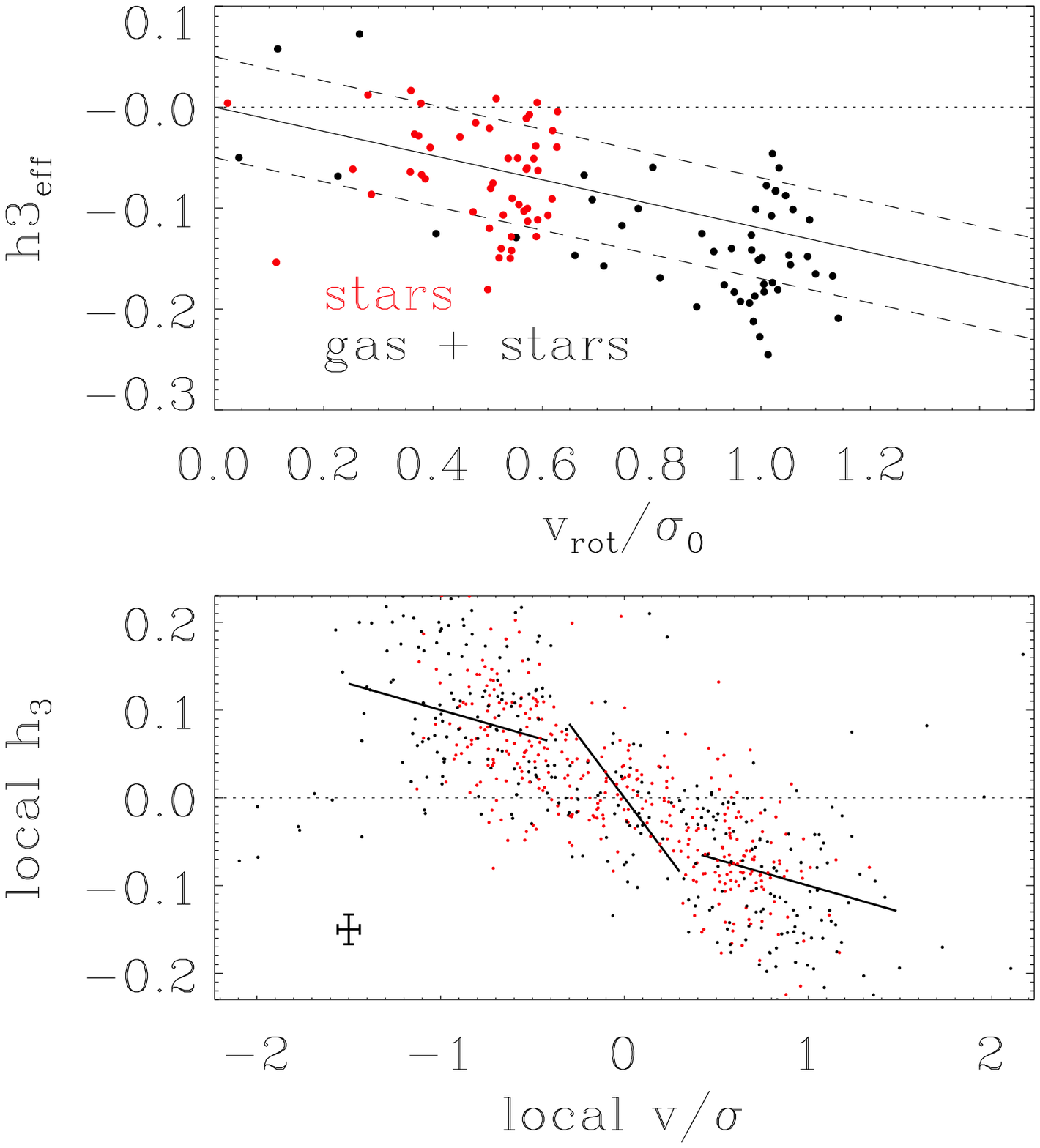}
\includegraphics{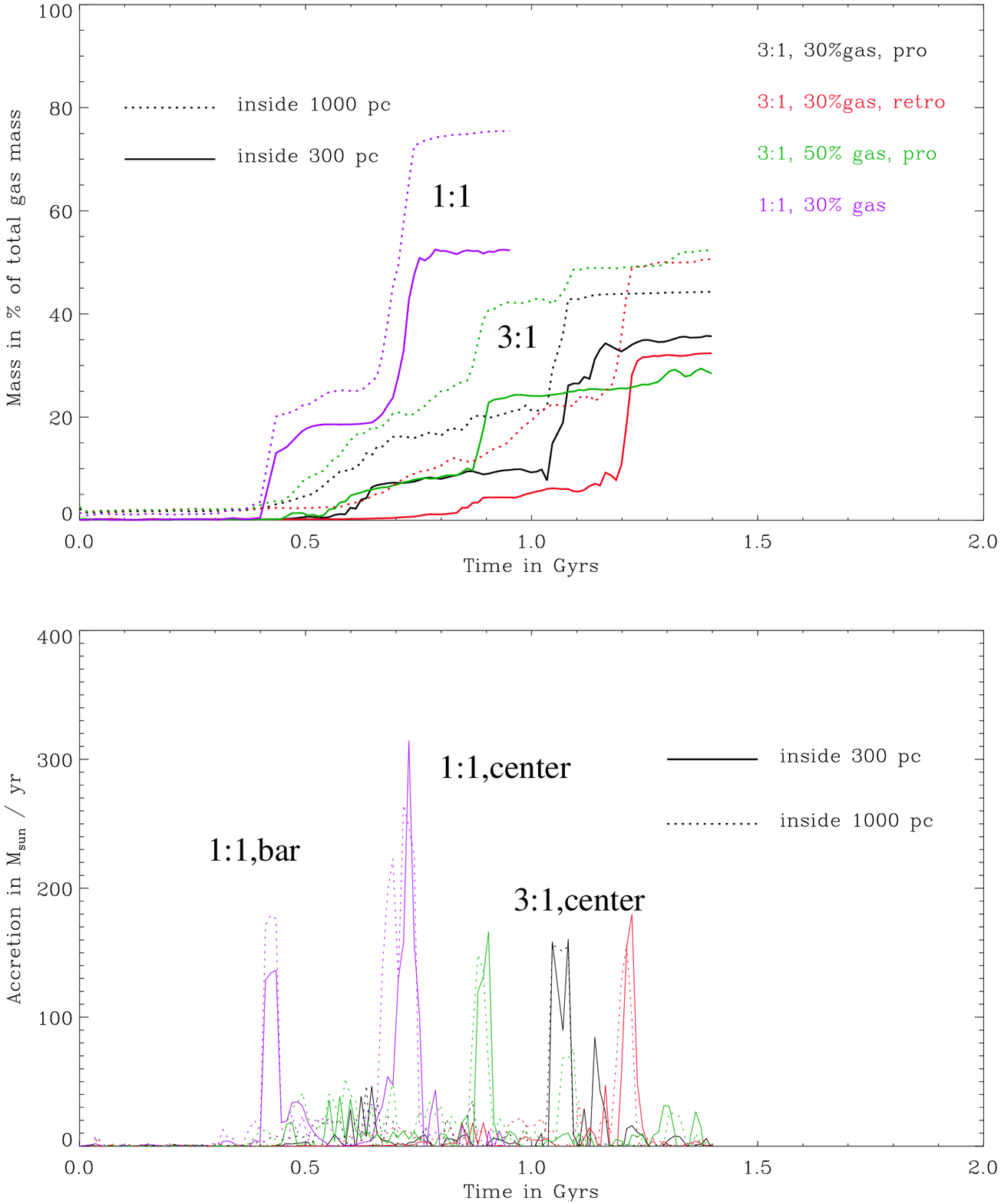}
\vspace{7cm}
\caption{{\it Right:} Global $h3_{eff}$ and local value of $h_3$ along
the line of sight with and without gas particles vs. rotational
support $v/\sigma$ for one 3:1 merger. Data from Bender et al. (1994)
are indicated by straight lines. {\it Left:} Upper diagram:
Mass inside 300pc and 1kpc vs. time for 4 simulations. Lower diagram:
Accretion rate during the bar formation and the merger of the central
parts of the galaxy onto the inner 300pc and 1kpc.}
\end{figure*}

The accretion history for one selected example of a 1:1 merger is shown in Figure
3. It can be divided in two major accretion events. First the gas  
flows to the center along a tidally induced bar and then the two
galaxy centers merge. In the end 50\% of the gas is funneled into the
central 300pc of the remnant. The centrifugal support of the gas in
3:1 mergers seems to be much stronger and prevents the gas from
clumping at the center. Instead the gas in unequal-mass mergers
settles in a large scale disk (Figure 2) and only around 30\% of the
gas is driven to the center, almost independent of the merger geometry
(Figure 3). The gas in the outer regions accumulates in dense knots
within tidal tails (see Figure 1: c and C) which could lead to the
formation of open clusters or dwarf satellites. Later on, the gas knots loose
angular momentum by dynamical friction and successively sink to the
center of the remnant thereby increasing the gas content of the disk. As in
the equal-mass case the main  accretion event is the merger of the two
galaxy centers.\\

The resulting stellar remnant has an oblate shape with
disky isophotes and kinematical properties in good agreement with
observations of faint ellipticals. In addition, we investigate the
LOSVD of one unequal-mass gas-rich merger 
remnant. Simulations of pure collisionless unequal-mass mergers
indicate a positive value of $h_3$, in contrast to observations (Naab \&
Burkert, 2001). In gas-rich mergers the sign of $h_3$ changes due to
the influence of the gas disk and is in good agreement with
observations of massive elliptical galaxies (Bender, Saglia \& Gerhard
1994). A comparison between simulated data and observations is shown
in Figure 3. The stellar particles alone and together with the gas
particles give the right correlation. Therefore the presence of
gas and the late formation of central disks might have played an
important role during the formation of elliptical galaxies.\\  

Simulations like this can help to understand the evolution of major
starbursts, AGNs and central power-law or disk-like structures that
are observed in fast rotating disky ellipticals (Faber et
al. 1997). We have to note that the influence of gas on the global  
structure of elliptical galaxies is not well understood since this
influence is sensitive to some certain details about star 
formation which is not included in the simulations we presented
here. Future investigations with better resolution (in combination
with GRAPE5) and carefully implemented recipes for starformation will
help to understand the role of gas dynamics and starformation in the
formation history of elliptical galaxies.

\end{document}